\documentclass[aps,prd,showpacs,showkeys,superscriptaddress,twocolumn]{revtex4-1}
\usepackage[colorlinks,linkcolor=blue,anchorcolor=blue,citecolor=blue,urlcolor=blue,breaklinks=true]{hyperref}
\usepackage{graphicx}
\usepackage{amsmath}
\usepackage{amssymb}
\usepackage{slashed}
\usepackage{latexsym}
\usepackage{epsfig}
\usepackage{amsbsy}
\usepackage{array}
\usepackage{changes}
\usepackage{amssymb}
\usepackage{setspace}
\usepackage{bm}
\usepackage{lipsum}
\usepackage{mathrsfs}
\usepackage{float}
\usepackage{color}
\usepackage{graphicx}
\usepackage{subfigure}
\usepackage[T1]{fontenc}
\usepackage{mathptmx}
\DeclareMathAlphabet{\mathcal}{OMS}{cmsy}{m}{n}
\DeclareSymbolFont{largesymbols}{OMX}{cmex}{m}{n}

\begin{document}
\author{Zheng Zhang}
\email{jozhzhang@163.com}
\affiliation{Department of physics, Nanjing University, Nanjing 210093, China}
\author{Chao Shi}
\email{shichao0820@gmail.com}
\affiliation{Department of nuclear science and technology,
Nanjing University of Aeronautics and Astronautics, Nanjing 210016, China}
\author{Hongshi Zong}
\email{zonghs@nju.edu.cn}
\affiliation{Department of physics, Nanjing University, Nanjing 210093, China}
\affiliation{Nanjing Proton Source Research and Design Center, Nanjing 210093, China}
\affiliation{Department of physics, Anhui Normal University, Wuhu, Anhui 241000, China }
\date{\today}

\title{Nambu-Jona-Lasinio model in a sphere}

\begin{abstract}
{We study the chiral phase transition of the two-flavor Nambu-Jona-Lasinio (NJL) model in a sphere with the MIT boundary condition. We find that the \textcolor{black}{spherical} MIT boundary condition results in  stronger finite size effects than the antiperiodic boundary condition. Our work may be helpful to study the finite size effects in heavy-ion collisions in a more realistic way.}
\bigskip


\end{abstract}

\maketitle


\maketitle

\section{Introduction}
\label{sec:Introduction}
The finite size effects in Quantum Chromodynamics (QCD) have caused much theoretical interest for more than two decades. The relavent study is important for the high-energy heavy ion collision (HIC) experiments. It is believed that these experiments could produces the quark-gluon plasma (QGP), a phase of matter believed to exist in the early universe. However, the QGP systems produced by HIC are always in a finite volume. For instance, the sizes of QGP are estimated to be between $2\ \mathrm{fm}$ to 10 fm in \cite{Palhares2011}, although the volume of Au-Au and Pb-Pb before freeze-out is about 50 $\mathrm{fm}^3$ to 250 $\mathrm{fm}^3$ \cite{Au2,Au3}. The finite size effects can modify the phase structure of strong interaction, dislocating critical lines and critical points, and also affect the dynamics of phase conversion \cite{finite1,finite2,Kiriyama2003,finite3,finite30,finite31,finite32,finite40,finite41,finite42,finite5,
Palhares2011,finite6,finite7,finite8}. The finite size effects have been investigated by different methods including chiral perturbation theory \cite{chiral2,chiral3}, quark-meson model \cite{finite40,finite42,quark-meson3,quark-meson4}, Dyson-Schwinger approach \cite{DS1,DS2,DS3,DS4}, Polyakov loop extended Nambu-Jona-Lasinio model \cite{pNJL1,pNJL2,pNJL3} and other non-perturbative renormalization group methods \cite{fluctuation}. A review of finite size effects can be found in \cite{review}. 

In most existing studies of finite size effects in QCD, the systems are usually treated as a box. However, we know that the QGP in HIC is more like a sphere than a box, so for a more realistic calculation, the shape effect should be considered. There are several studies \cite{Kiriyama2003,finite3,MRE2,MRE3,MRE4} treating the system as a sphere, and the multiple reflection expansion (MRE) method \cite{mre} is used.  But the MRE method is essentially an asymptotic expansion method which becomes invalid for very small volume.
 So we want to find a better method to deal with the finite size effects in a sphere, which is one of the major subjects of this work.

\textcolor{black}{To give a brief introduction to our method, let's first recall how we deal with finite size effects in a box.}  Usually, we put the system into a box and a spatial boundary condition is required, which results in discretized momenta in the spatial direction. To consider finite size effects, we replace the integral over spatial momenta with a sum over discrete momentum modes. This "brute force method" has no difficulty to be applied for the sphere case, though the calculation is more complicated. In this work, we will use this method to study the chiral phase transition of NJL model in a sphere.

This paper is organized as follows: In Section 2, The gap equation of NJL model in a sphere with the MIT boundary condition is derived. The chiral phase transition of the model is presented in Section 3, and a summary is given in Section 4.

\section{NJL model in a sphere with the MIT boundary condition}
\label{sec:NJL model}
NJL model is a low energy effective theory of QCD. It has the feature of dynamical chiral symmetry breaking. The Lagrangian of the two-flavor NJL model is
\begin{equation}
\mathscr{L}=\overline{\psi}(i\gamma^\mu\partial_\mu-m)\psi+G[(\overline{\psi}\psi)^2+(\overline{\psi}i\gamma^5\tau\psi)^2],
\end{equation}
where $G$ is the effective coupling, and $m$ is the current quark mass. Here we consider $u$ and $d$ quarks with exact isospin symmetry.
In the mean field approximation, the gap equation is given as (only consider the Hartree term)  \cite{NJLreview} 
\begin{equation}
M=m-2G\left\langle\overline{\psi}\psi \right\rangle,
\end{equation}
At zero temperature, the condensation
\begin{equation}\label{cond}
\left\langle\overline{\psi}\psi \right\rangle=i N_cN_f\int\frac{d^4p}{(2\pi)^4}\mathrm{tr}S(p),
\end{equation}
where $N_c=3$ is the number of colors, $N_f=2$ is the number of flavors, and
\begin{equation}
S(p)=\frac{\not{p}+M}{p^2-M^2}.
\end{equation}
The trace is taken over the Dirac indices. Since the NJL model is non-renormalizable, regularization is needed. A widely used regularization scheme is the three momentum cut-off regularization \cite{NJLreview}, which ingores the high-frequency modes. We think it is not very suitable for dealing with finite size effects because in a finite volume only high-frequecy modes can exist. So, in this paper we adopt the proper time regularization \cite{NJLreview}, which takes into account the contribution of all the modes. With this regularization, the condensation in the infinite volume at zero temperature can be written as
\begin{equation}\label{gapeq}
\begin{aligned}
\left\langle\overline{\psi}\psi \right\rangle&=-4N_cN_fM\int\frac{d^4p}{(2\pi)^4}\int_{\tau_{UV}}^\infty \mathrm{d\tau}e^{-\tau(p^2+M^2)}\\
&=-\frac{3M}{2\pi^2}\int_{\tau_{UV}}^{\infty}\mathrm{d\tau}\frac{e^{-\tau M^2}}{\tau^2}.
\end{aligned}
\end{equation}
 \textcolor{black}{At finite temperature, the integration of $p_0$ is replaced by a sum of all fermion Matsubara frequencies $\omega_k=(2k+1)\pi T$. Then the condensation at finite temperature can be written as}
\begin{equation}\label{finitec}
\begin{aligned}
\left\langle\overline{\psi}\psi \right\rangle&=-4N_cN_fM\int_{\tau_{UV}}^\infty \mathrm{d\tau}e^{-\tau M^2}\times T\sum_{k=-\infty}^{\infty}\int\frac{d^3\vec{p}}{(2\pi)^3}e^{-\tau (p^2+\omega_k^2)}\\
&=-\frac{3MT}{\pi^{3/2}}\int_{\tau_{UV}}^{\infty}\mathrm{d\tau}\frac{e^{-\tau M^2}}{\tau^{3/2}}\theta_2(0,e^{-4\pi^2\tau T^2}).
\end{aligned}
\end{equation}
\textcolor{black}{where $\theta_2(0,q)=2\sqrt[4]{q}\sum_{n=0}^{\infty}q^{n(n+1)}$. Then the gap equation at finite temperature is}
\begin{equation}\label{gapff}
M=m+\frac{6GMT}{\pi^{3/2}}\int_{\tau_{UV}}^{\infty}\mathrm{d\tau}\frac{e^{-\tau M^2}}{\tau^{3/2}}\theta_2(0,e^{-4\pi^2\tau T^2}).
\end{equation}

 Equation (\ref{gapff}) is the usual gap equation of NJL model in infinite volume. For finite size system, it should be modified. As we stated in the Introduction, three momenta will be discretized by the boundary condition. For example, the allowed values of momentum modes in a box under antiperiodic boundary condition are 
\begin{equation}
\vec{p}_{\mathrm{APBC}}^2=\frac{4\pi^2}{L^2}\sum_{i=1}^{3}(n_i+\frac{1}{2})^2,\ \ n_i=0,\pm1,\pm2,....
\end{equation}
To modify the gap equation, we replace the integral over momentum with a sum over all allowed momentum modes, i.e.
\begin{equation}
\int \frac{d^3\vec{p}}{(2\pi)^3} \to \frac{1}{V}\sum_{p_k}.
\end{equation}
\textcolor{black}{where $p_k$ stands for $|\vec{p_k}|$.} This replacement has been used in many works, e.g. \cite{Palhares2011,fluctuation,DS4,qingwu}. \textcolor{black}{Here we note that the finite size effects are not fully accounted for by the replacement of the intergral by sum over discrete modes. The condensation is inhomogeneous in a finite system in general. For simplicity, we negelect the inhomogeneous effects and treat $\left\langle\overline{\psi}\psi \right\rangle$ as a constant here. The modified gap equation in finite volume under antiperiodic boundary condition at finite temperature is }
\begin{equation}\label{gapf}
M=m+\frac{48GMT}{V}\int_{\tau_{UV}}^{\infty}\mathrm{d\tau}e^{-\tau M^2}\theta_2(0,e^{-4\pi^2\tau T^2})\sum_{p_k}e^{-\tau p_k^2}.
\end{equation}
Here we want to note that the above method to deal with the finite size effects in NJL model and the regularization scheme are the same as that in \cite{qingwu}. There are other methods and regularization schemes employed in different studies \cite{finite3,finite30,Xia}.

Now we want to follow the above procedure to get the gap equation of NJL model in a sphere. First, we should select a proper boundary condition. In the past studies of finite size effects, periodic and antiperiodic boundary conditions are usually selected. However, it's hard to define periodic and antiperiodic on a sphere. Dirichlet boundary condition can be defined on a sphere, but for fermions,  it is so strict that no solution exists \cite{Diraceq}. Here we select the MIT boundary condition, first proposed in the MIT bag model \cite{MIT1,MIT2}. For a sphere, it can be written as
\begin{equation}\label{bc}
-i\hat{r}\cdot\vec{\gamma}\psi(t,r,\theta,\phi)|_{r=R}=\psi(t,r,\theta,\phi)|_{r=R},
\end{equation}
where $\hat{r}$ is the unit vector normal to the sphere surface, and $\vec{\gamma}=(\gamma^1,\gamma^2,\gamma^3)$. This boundary condition confines the fermions inside the cavity since it forces the normal component of the fermionic current $\overline{\psi}\gamma^\mu\psi$ to be zero at the surface of the cavity \cite{MIT1}. We think the MIT boundary condition is more suitable for the finite size effects study in HICs than (anti)periodic boundary conditions for its confinement character. Note the authors of \cite{cylinder} already investigated the NJL model in a cylinder with the MIT boundary condition. Here we  investigate the NJL model in a sphere with also the MIT boundary condition.

Once the boundary condition Eq. (\ref{bc}) has been selected, the discrete values of the momentum can be obtained by solving the equation of motion of the NJL model with Eq. (\ref{bc}). The NJL model after mean field approximation can be seen as a model of free particle with mass $M$, and its equation of motion is the free Dirac equation. Under spherical MIT boundary condition, the allowed momentum values are given by the following eigen-equations \cite{Greiner}
\begin{equation}\label{engin}
j_{l_\kappa}(pR)=-\operatorname{sgn}(\kappa)\frac{p}{E+M}j_{\overline{l}_\kappa}(pR),
\end{equation}
where $$l_{\kappa}=\left\{\begin{array}{cl}{-\kappa-1} & {\text { for } \kappa<0} \\ {\kappa} & {\text { for } \kappa>0}\end{array}\right.,$$
$$\overline{l}_{\kappa}=\left\{\begin{array}{cc}{-\kappa} & {\text { for } \kappa<0} \\ {\kappa-1} & {\text { for } \kappa>0}\end{array}\right.,$$
 $\kappa=\pm1,\pm2,...$ and $j_l(x)$ is the $l$-th ordered spherical Bessel function.  The $p$ stands for $|\vec{p}|$.
With the following replacement
\begin{equation}\label{rep}
\int \frac{d^3\vec{p}}{(2\pi)^3} \to \frac{1}{2V}(\sum_{p_k,\kappa>0}+\sum_{p_k,\kappa<0}),
\end{equation}
We can get the gap equation of NJL model in a sphere with the MIT boundary condition. The factor 2 in Eq. (\ref{rep}) comes from the nondegeneracy of $\kappa$ and $-\kappa$ states. The final gap equation in a sphere at finite temperature is
\begin{equation}\label{gapfb}
\begin{aligned}
M=&m+\frac{24GMT}{V}\int_{\tau_{UV}}^{\infty}\mathrm {d\tau}e^{-\tau M^2}\theta_2(0,e^{-4\pi^2\tau T^2})\times \\ &(\sum_{p_k,\kappa>0}e^{-\tau p_k^2}+\sum_{p_k,\kappa<0}e^{-\tau p_k^2}).
\end{aligned}
\end{equation}

\section{Chiral phase transition of the NJL model in a sphere}
\label{Result}

\begin{figure}
\begin{minipage}[b]{0.45\textwidth}
\includegraphics[width=1\linewidth]{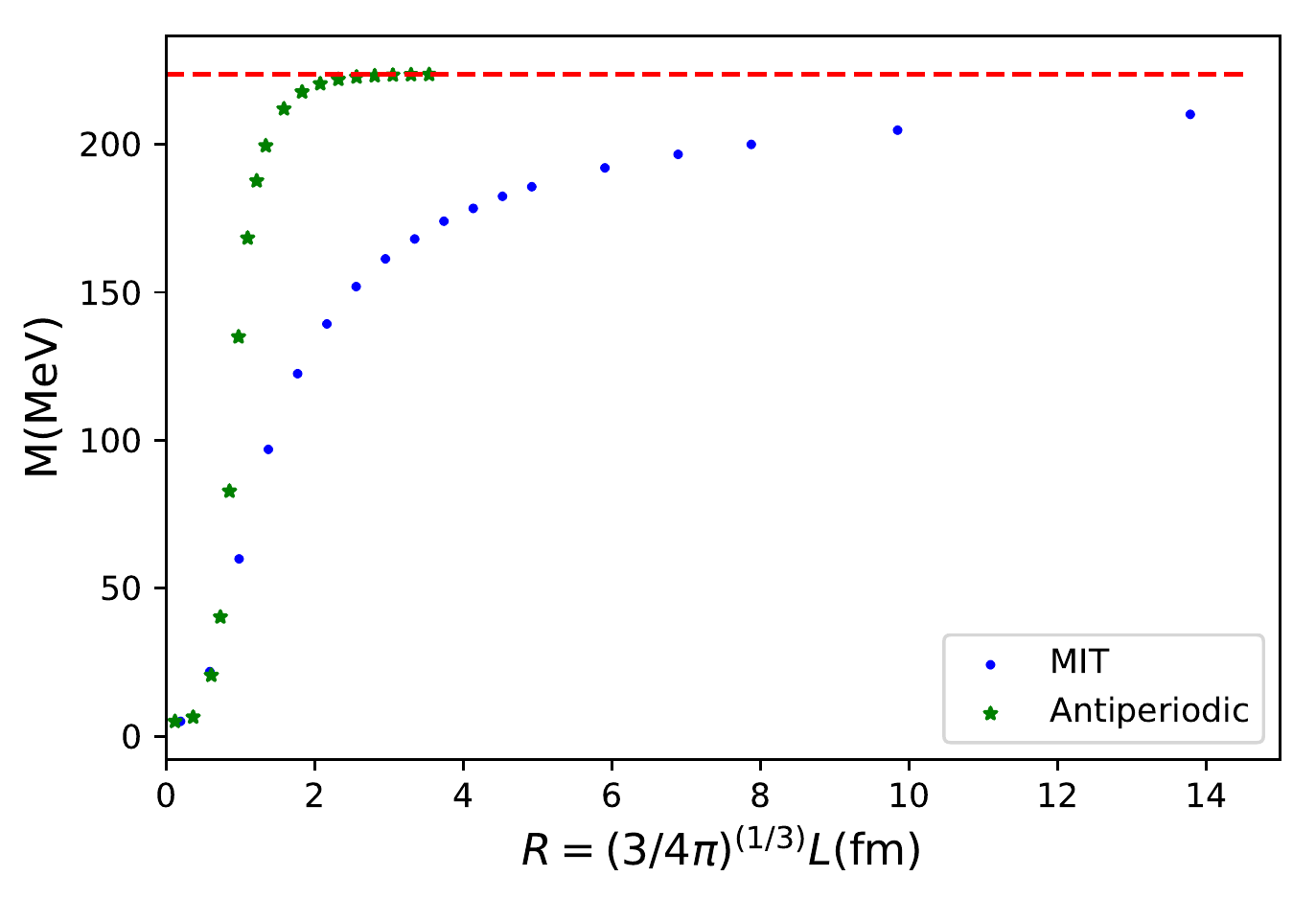} \\
\end{minipage}
\caption{(Color online) Constituent quark mass $M$ as a function of radius $R$ and box size $L$ at zero temperature. We set the x-axis to be $R$ for the MIT boundary condition case and $({3}/{4\pi})^{{1}/{3}}L$ for the antiperiodic boundary condition case so that they have the same volume. The red dashed line is the constituent quark mass in infinite volume.}
\label{zerot}
\end{figure}

Solving the modified gap equation Eq. (\ref{gapfb}) at zero temperature with different radii, we can find how the finite size influences the constituent quark mass at zero temperature. It is well known that spontaneous symmetry breaking can only occur in infinitely large systems in principle \cite{Leut,Weinberg:1996kr}. So in our cases, we expect the small size will lead to the restoration of chiral symmetry. In fact, the decrease of the volume has a similar effect to the increase of the temperature, so we expect the constituent quark mass will decrease when the volume decreases. Note these statements have been confirmed in \cite{review, qingwu, Xia}. 
Figure \ref{zerot} presents the constituent quark mass $M$ as a function of the radius $R$ at zero temperature. The result under antiperiodic boundary condition is also presented for comparison. The parameters are chosen as $m=5\ \mathrm{MeV}, \tau_{UV}=1/1080^2\ \mathrm{MeV^{-2}}, G=3.26\times10^{-6} \mathrm{MeV^{-2}}$. For the spherical MIT boundary condition case, we find when the radius reaches about \textcolor{black}{14 fm}, the constituent quark mass gets very close to that in an infinite system. That is to say, a system whose size is above \textcolor{black}{14 fm} can be regarded as an infinitely large system.  \textcolor{black}{But for antiperiodic boundary condition case, when $L>3$ fm, the box is already large enough to be regarded as an infinite system. We think the stronger finite size effects induced by the spherical MIT boundary condition may be due to the confinement character of this boundary condition.}
\begin{figure}[t]
\begin{minipage}[t]{0.47\textwidth}
\includegraphics[width=1\linewidth]{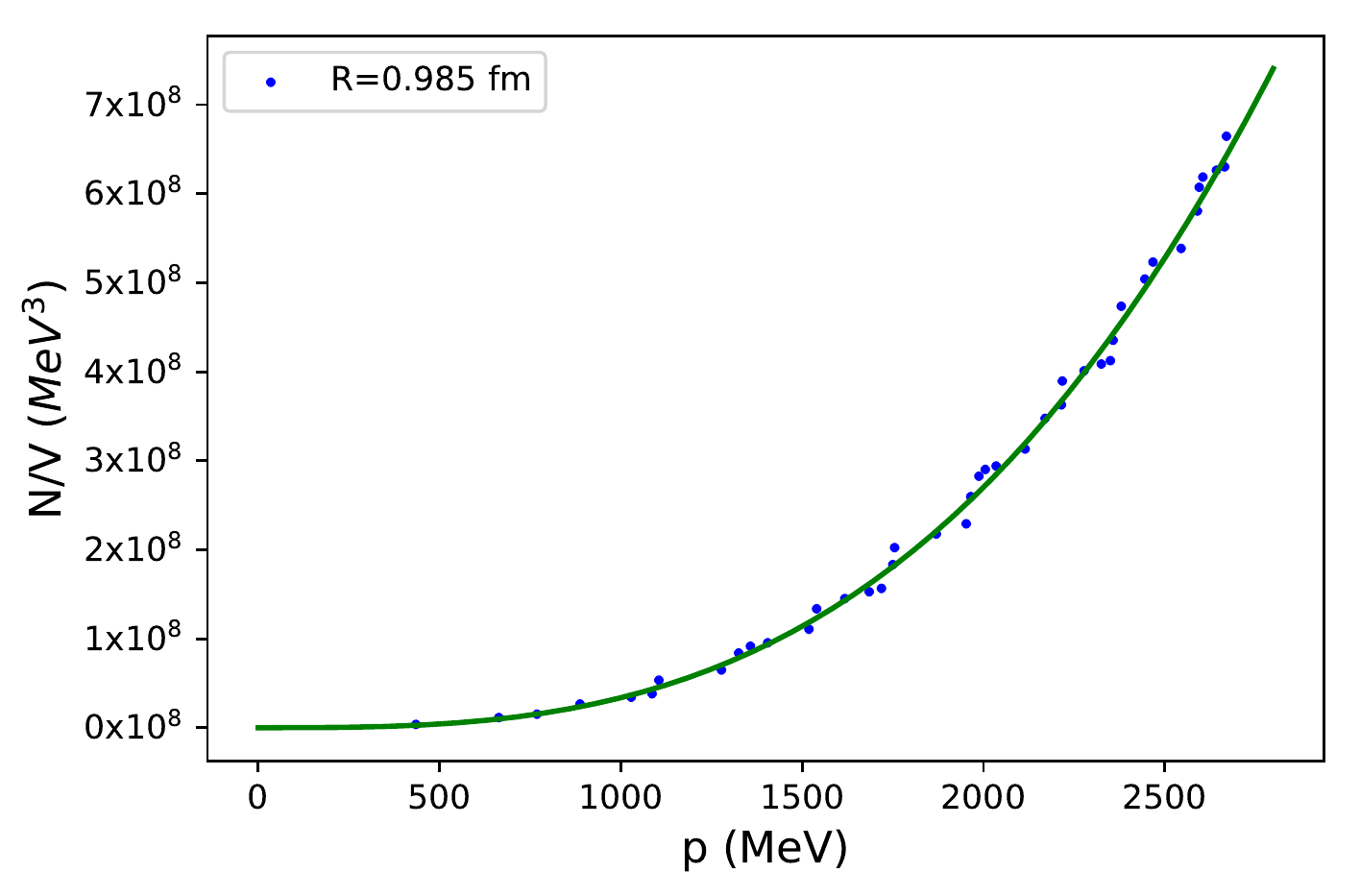} \\
\end{minipage}

\begin{minipage}[t]{0.47\textwidth}
\includegraphics[width=1\linewidth]{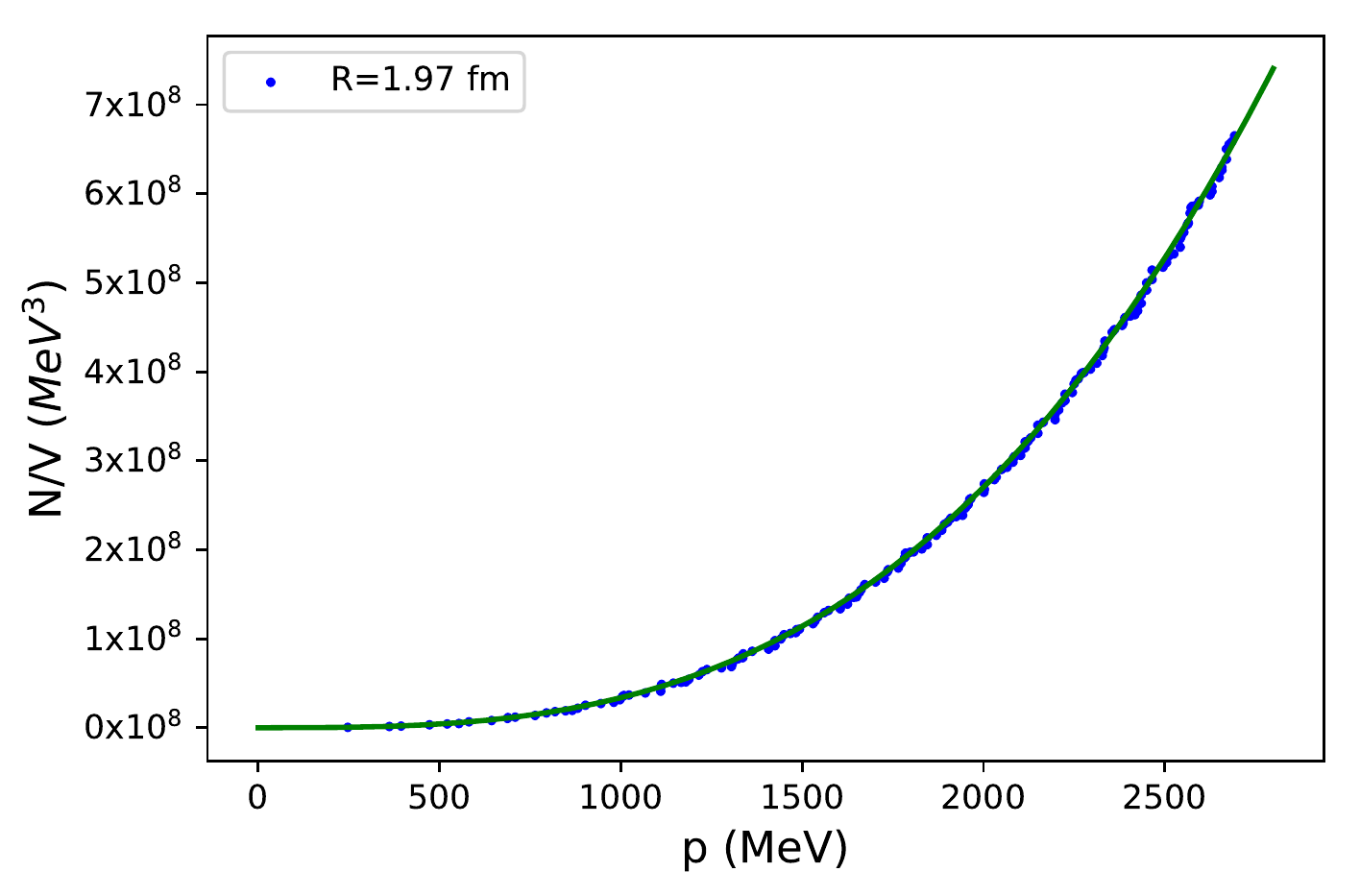} \\
\end{minipage}
\caption{(Color online) The integrated number of modes as a function of the single particle momentum of quarks per volumn, calculated under spherical MIT boundary condition with $R=0.985$ fm and $R=1.97$ fm at zero temperature. The solid line is the infinite volume limit.}
\label{nmodes}
\end{figure}

\textcolor{black}{We know the number of modes of quarks is essential to cause the spontaneous breaking of the chiral symmetry. In Fig. \ref{nmodes}, we show the integrated number of modes as a function of the single particle momentum of quarks per volume in finite volume, and the infinite limit is shown for comparison. We can observe that the integrated number of modes at finite volume fluctuates around the large volume limit.}

\begin{figure}[h]
\begin{minipage}[t]{0.47\textwidth}
\includegraphics[width=1\linewidth]{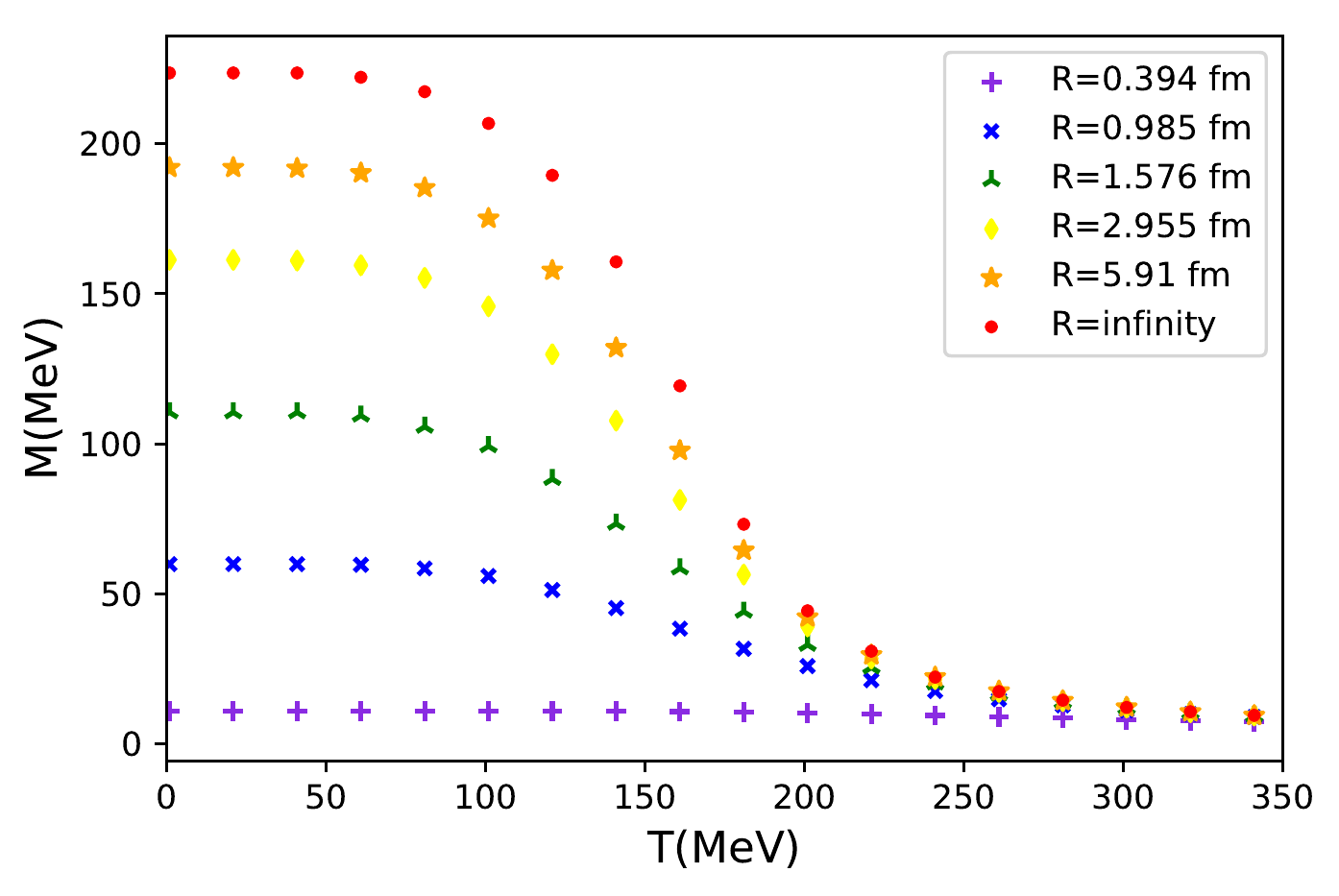} \\
\end{minipage}
\caption{(Color online) Constituent quark mass $M$ as functions of temperature in spheres with different radii.}
\label{fint}
\end{figure}

\begin{figure}[h]
\begin{minipage}[t]{0.47\textwidth}
\includegraphics[width=1\linewidth]{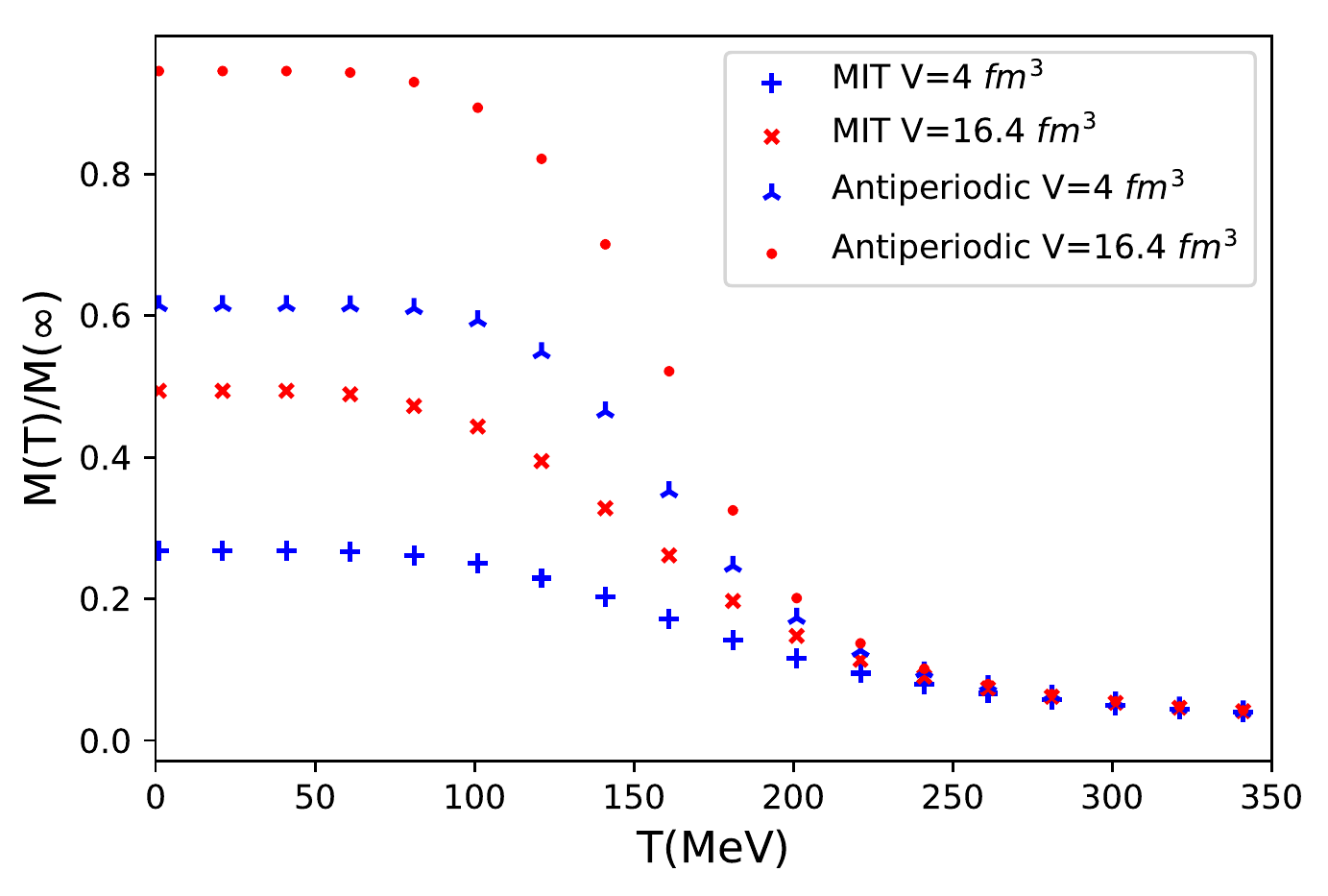} \\
\end{minipage}
\caption{(Color online) $M(T)/M(\infty)$ under MIT boundary condition and antiperiodic boundary condition.}
\label{mt}
\end{figure}

\begin{figure}[h]
\begin{minipage}[t]{0.47\textwidth}
\includegraphics[width=1\linewidth]{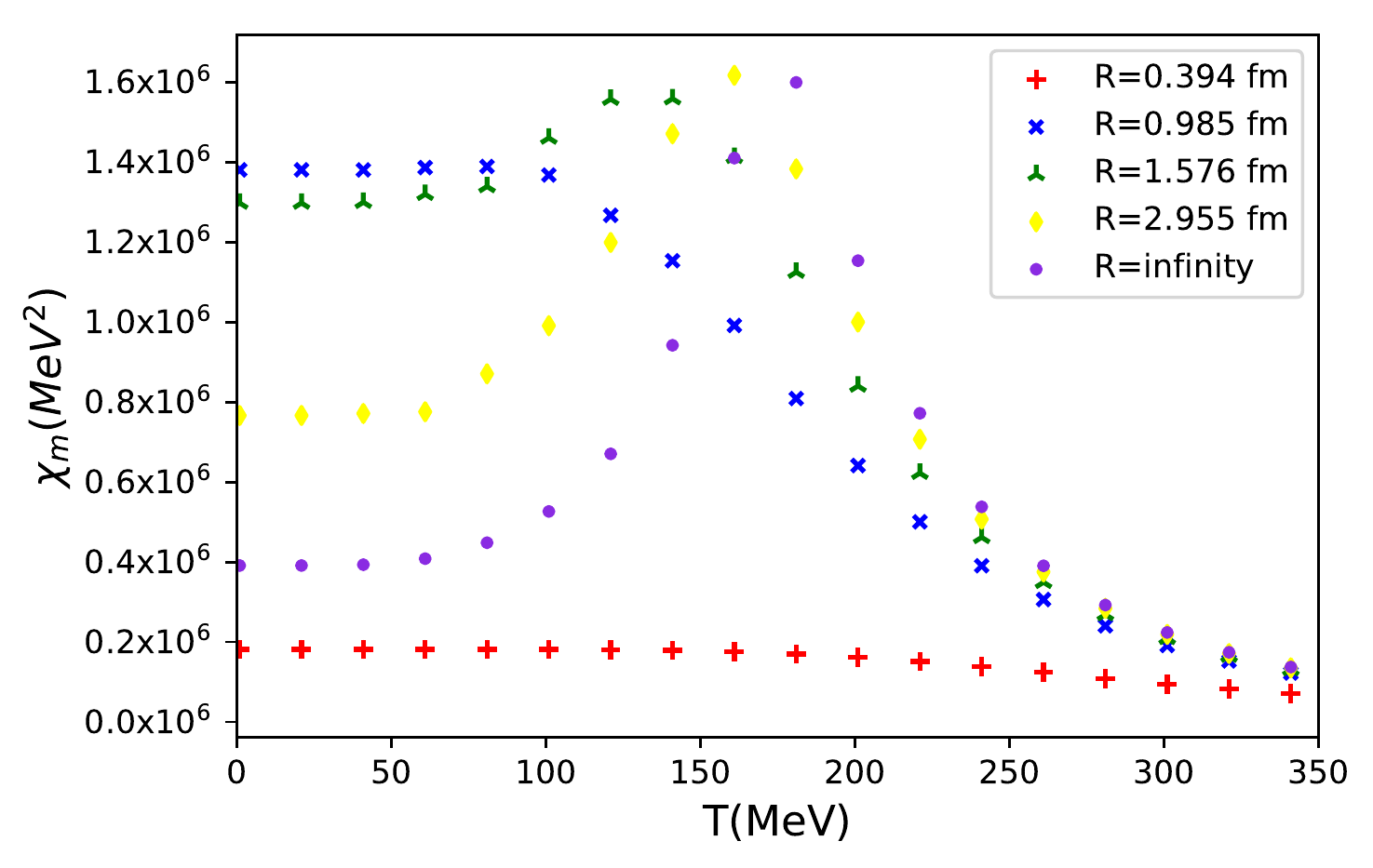} \\
\end{minipage}
\caption{(Color online) Chiral susceptibility under MIT boundary condition with different radii.}
\label{sus}
\end{figure}

To show the influence of the finite size on the chiral phase transition of the NJL model, we solve the gap equation Eq. (\ref{gapfb}) at finite temperature with different radii. Figure \ref{fint} presents the constituent masses as functions of temperature in spheres with different radii. We find at the same temperature, the constituent quark mass in small volume is smaller than that in large volume because the finite size partially restores the chiral symmetry. \textcolor{black}{ Figure \ref{mt} compares the results under the spherical MIT boundary condition and antiperiodic boundary condition at finite temperature, which also shows the spherical MIT boundary condition leads to stronger finite size effect than antiperiodic boundary condition. Figure \ref{sus} shows the chiral susceptibility $\chi_m=-\partial\left\langle\overline{\psi}\psi \right\rangle/\partial m$ under MIT boundary condition, we find the finite size smoothes the peaks of the chiral susceptibility, which is a well-known finite size effect.}

\textcolor{black}{Since the sizes of QGP systems are estimated to be between 2 fm to 10 fm according to  \cite{Palhares2011}, our results indicate these systems have considerable finite size effects. In some previous work in which periodic or antiperiodic boundary condition is adopted, e.g., \cite{finite40,Xia}, the finite size effects are weaker than our results, which is consistent with our finding in this paper that antiperiodic boundary condition leads to weaker finite size effects than spherical MIT boundary condition. Thus, we want to call attention to the importance of boundary conditions when studying finite size effects. In fact, the importance of boundary conditions had been noticed by some previous work \cite{finite2}. }

\section{Summary}
\label{sec:Summary}

In this work, we apply the widely used "brute force method" to study the finite size effects of the NJL model in a sphere with the MIT boundary condition. The chiral phase transition is investigated in this model. We find the spherical MIT boundary condition leads to stronger finite size effects than antiperiodic boundary condition. Since we believe the systems we deal with here are more close to the QGP systems in HIC, we think the finite size effects in HIC may be stronger than former estimations, so it deserves careful study in the future.

\section*{Acknowledgements}
This work is supported in part by the National Natural Science Foundation of China (under Grants No. 11475085, No. 11535005, No. 11905104 and No. 11690030) and by Nation Major State Basic Research and Development of China (2016YFE0129300). 

\bibliography{ref}

\begin{thebibliography}{44}%
\makeatletter
\providecommand \@ifxundefined [1]{%
 \@ifx{#1\undefined}
}%
\providecommand \@ifnum [1]{%
 \ifnum #1\expandafter \@firstoftwo
 \else \expandafter \@secondoftwo
 \fi
}%
\providecommand \@ifx [1]{%
 \ifx #1\expandafter \@firstoftwo
 \else \expandafter \@secondoftwo
 \fi
}%
\providecommand \natexlab [1]{#1}%
\providecommand \enquote  [1]{``#1''}%
\providecommand \bibnamefont  [1]{#1}%
\providecommand \bibfnamefont [1]{#1}%
\providecommand \citenamefont [1]{#1}%
\providecommand \href@noop [0]{\@secondoftwo}%
\providecommand \href [0]{\begingroup \@sanitize@url \@href}%
\providecommand \@href[1]{\@@startlink{#1}\@@href}%
\providecommand \@@href[1]{\endgroup#1\@@endlink}%
\providecommand \@sanitize@url [0]{\catcode `\\12\catcode `\$12\catcode
  `\&12\catcode `\#12\catcode `\^12\catcode `\_12\catcode `\%12\relax}%
\providecommand \@@startlink[1]{}%
\providecommand \@@endlink[0]{}%
\providecommand \url  [0]{\begingroup\@sanitize@url \@url }%
\providecommand \@url [1]{\endgroup\@href {#1}{\urlprefix }}%
\providecommand \urlprefix  [0]{URL }%
\providecommand \Eprint [0]{\href }%
\providecommand \doibase [0]{http://dx.doi.org/}%
\providecommand \selectlanguage [0]{\@gobble}%
\providecommand \bibinfo  [0]{\@secondoftwo}%
\providecommand \bibfield  [0]{\@secondoftwo}%
\providecommand \translation [1]{[#1]}%
\providecommand \BibitemOpen [0]{}%
\providecommand \bibitemStop [0]{}%
\providecommand \bibitemNoStop [0]{.\EOS\space}%
\providecommand \EOS [0]{\spacefactor3000\relax}%
\providecommand \BibitemShut  [1]{\csname bibitem#1\endcsname}%
\let\auto@bib@innerbib\@empty
\bibitem [{\citenamefont {Palhares}\ \emph {et~al.}(2011)\citenamefont
  {Palhares}, \citenamefont {Fraga},\ and\ \citenamefont
  {Kodama}}]{Palhares2011}%
  \BibitemOpen
  \bibfield  {author} {\bibinfo {author} {\bibfnamefont {L.~F.}\ \bibnamefont
  {Palhares}}, \bibinfo {author} {\bibfnamefont {E.~S.}\ \bibnamefont {Fraga}},
  \ and\ \bibinfo {author} {\bibfnamefont {T.}~\bibnamefont {Kodama}},\ }\href
  {\doibase 10.1088/0954-3899/38/8/085101} {\bibfield  {journal} {\bibinfo
  {journal} {J. Phys. G: Nucl. Part. Phys.}\ }\textbf {\bibinfo {volume}
  {38}},\ \bibinfo {pages} {085101} (\bibinfo {year} {2011})}\BibitemShut
  {NoStop}%
\bibitem [{\citenamefont {Bass}\ \emph {et~al.}(1999)\citenamefont {Bass},
  \citenamefont {Weber}, \citenamefont {Ernst}, \citenamefont {Bleicher},
  \citenamefont {Belkacem}, \citenamefont {Bravina}, \citenamefont {Soff},
  \citenamefont {Stöcker}, \citenamefont {Greiner},\ and\ \citenamefont
  {Spieles}}]{Au2}%
  \BibitemOpen
  \bibfield  {author} {\bibinfo {author} {\bibfnamefont {S.}~\bibnamefont
  {Bass}}, \bibinfo {author} {\bibfnamefont {H.}~\bibnamefont {Weber}},
  \bibinfo {author} {\bibfnamefont {C.}~\bibnamefont {Ernst}}, \bibinfo
  {author} {\bibfnamefont {M.}~\bibnamefont {Bleicher}}, \bibinfo {author}
  {\bibfnamefont {M.}~\bibnamefont {Belkacem}}, \bibinfo {author}
  {\bibfnamefont {L.}~\bibnamefont {Bravina}}, \bibinfo {author} {\bibfnamefont
  {S.}~\bibnamefont {Soff}}, \bibinfo {author} {\bibfnamefont {H.}~\bibnamefont
  {Stöcker}}, \bibinfo {author} {\bibfnamefont {W.}~\bibnamefont {Greiner}}, \
  and\ \bibinfo {author} {\bibfnamefont {C.}~\bibnamefont {Spieles}},\ }\href
  {\doibase https://doi.org/10.1016/S0146-6410(99)00086-1} {\bibfield
  {journal} {\bibinfo  {journal} {Prog. Part. Nucl. Phys.}\ }\textbf {\bibinfo
  {volume} {42}},\ \bibinfo {pages} {313 } (\bibinfo {year}
  {1999})}\BibitemShut {NoStop}%
\bibitem [{\citenamefont {Gr\"af}\ \emph {et~al.}(2012)\citenamefont {Gr\"af},
  \citenamefont {Bleicher},\ and\ \citenamefont {Li}}]{Au3}%
  \BibitemOpen
  \bibfield  {author} {\bibinfo {author} {\bibfnamefont {G.}~\bibnamefont
  {Gr\"af}}, \bibinfo {author} {\bibfnamefont {M.}~\bibnamefont {Bleicher}}, \
  and\ \bibinfo {author} {\bibfnamefont {Q.}~\bibnamefont {Li}},\ }\href
  {\doibase 10.1103/PhysRevC.85.044901} {\bibfield  {journal} {\bibinfo
  {journal} {Phys. Rev. C}\ }\textbf {\bibinfo {volume} {85}},\ \bibinfo
  {pages} {044901} (\bibinfo {year} {2012})}\BibitemShut {NoStop}%
\bibitem [{\citenamefont {Gopie}\ and\ \citenamefont
  {Ogilvie}(1999)}]{finite1}%
  \BibitemOpen
  \bibfield  {author} {\bibinfo {author} {\bibfnamefont {A.}~\bibnamefont
  {Gopie}}\ and\ \bibinfo {author} {\bibfnamefont {M.~C.}\ \bibnamefont
  {Ogilvie}},\ }\href {\doibase 10.1103/PhysRevD.59.034009} {\bibfield
  {journal} {\bibinfo  {journal} {Phys. Rev. D}\ }\textbf {\bibinfo {volume}
  {59}},\ \bibinfo {pages} {034009} (\bibinfo {year} {1999})}\BibitemShut
  {NoStop}%
\bibitem [{\citenamefont {Bazavov}\ and\ \citenamefont {Berg}(2007)}]{finite2}%
  \BibitemOpen
  \bibfield  {author} {\bibinfo {author} {\bibfnamefont {A.}~\bibnamefont
  {Bazavov}}\ and\ \bibinfo {author} {\bibfnamefont {B.~A.}\ \bibnamefont
  {Berg}},\ }\href {\doibase 10.1103/PhysRevD.76.014502} {\bibfield  {journal}
  {\bibinfo  {journal} {Phys. Rev. D}\ }\textbf {\bibinfo {volume} {76}},\
  \bibinfo {pages} {014502} (\bibinfo {year} {2007})}\BibitemShut {NoStop}%
\bibitem [{\citenamefont {Kiriyama}\ and\ \citenamefont
  {Hosaka}(2003)}]{Kiriyama2003}%
  \BibitemOpen
  \bibfield  {author} {\bibinfo {author} {\bibfnamefont {O.}~\bibnamefont
  {Kiriyama}}\ and\ \bibinfo {author} {\bibfnamefont {A.}~\bibnamefont
  {Hosaka}},\ }\href {\doibase 10.1103/PhysRevD.67.085010} {\bibfield
  {journal} {\bibinfo  {journal} {Phys. Rev. D}\ }\textbf {\bibinfo {volume}
  {67}},\ \bibinfo {pages} {085010} (\bibinfo {year} {2003})}\BibitemShut
  {NoStop}%
\bibitem [{\citenamefont {Kiriyama}\ \emph {et~al.}()\citenamefont {Kiriyama},
  \citenamefont {Kodama},\ and\ \citenamefont {Koide}}]{finite3}%
  \BibitemOpen
  \bibfield  {author} {\bibinfo {author} {\bibfnamefont {O.}~\bibnamefont
  {Kiriyama}}, \bibinfo {author} {\bibfnamefont {T.}~\bibnamefont {Kodama}}, \
  and\ \bibinfo {author} {\bibfnamefont {T.}~\bibnamefont {Koide}},\ }\href
  {https://arxiv.org/abs/hep-ph/0602086} {\bibinfo  {journal}
  {arXiv:hep-ph/0602086}\ }\BibitemShut {NoStop}%
\bibitem [{\citenamefont {Abreu}\ \emph {et~al.}(2006)\citenamefont {Abreu},
  \citenamefont {Gomes},\ and\ \citenamefont {da~Silva}}]{finite30}%
  \BibitemOpen
\bibfield  {journal} {  }\bibfield  {author} {\bibinfo {author} {\bibfnamefont
  {L.}~\bibnamefont {Abreu}}, \bibinfo {author} {\bibfnamefont
  {M.}~\bibnamefont {Gomes}}, \ and\ \bibinfo {author} {\bibfnamefont
  {A.}~\bibnamefont {da~Silva}},\ }\href {\doibase
  https://doi.org/10.1016/j.physletb.2006.10.015} {\bibfield  {journal}
  {\bibinfo  {journal} {Phys. Lett. B}\ }\textbf {\bibinfo {volume} {642}},\
  \bibinfo {pages} {551 } (\bibinfo {year} {2006})}\BibitemShut {NoStop}%
\bibitem [{\citenamefont {Abreu}\ \emph {et~al.}(2009)\citenamefont {Abreu},
  \citenamefont {Malbouisson}, \citenamefont {Malbouisson},\ and\ \citenamefont
  {Santana}}]{finite31}%
  \BibitemOpen
  \bibfield  {author} {\bibinfo {author} {\bibfnamefont {L.}~\bibnamefont
  {Abreu}}, \bibinfo {author} {\bibfnamefont {A.}~\bibnamefont {Malbouisson}},
  \bibinfo {author} {\bibfnamefont {J.}~\bibnamefont {Malbouisson}}, \ and\
  \bibinfo {author} {\bibfnamefont {A.}~\bibnamefont {Santana}},\ }\href
  {\doibase https://doi.org/10.1016/j.nuclphysb.2009.04.012} {\bibfield
  {journal} {\bibinfo  {journal} {Nucl. Phys. B}\ }\textbf {\bibinfo {volume}
  {819}},\ \bibinfo {pages} {127 } (\bibinfo {year} {2009})}\BibitemShut
  {NoStop}%
\bibitem [{\citenamefont {Abreu}\ \emph {et~al.}(2011)\citenamefont {Abreu},
  \citenamefont {Malbouisson},\ and\ \citenamefont {Malbouisson}}]{finite32}%
  \BibitemOpen
  \bibfield  {author} {\bibinfo {author} {\bibfnamefont {L.~M.}\ \bibnamefont
  {Abreu}}, \bibinfo {author} {\bibfnamefont {A.~P.~C.}\ \bibnamefont
  {Malbouisson}}, \ and\ \bibinfo {author} {\bibfnamefont {J.~M.~C.}\
  \bibnamefont {Malbouisson}},\ }\href {\doibase 10.1103/PhysRevD.83.025001}
  {\bibfield  {journal} {\bibinfo  {journal} {Phys. Rev. D}\ }\textbf {\bibinfo
  {volume} {83}},\ \bibinfo {pages} {025001} (\bibinfo {year}
  {2011})}\BibitemShut {NoStop}%
\bibitem [{\citenamefont {Braun}\ \emph
  {et~al.}(2005{\natexlab{a}})\citenamefont {Braun}, \citenamefont {Klein},\
  and\ \citenamefont {Pirner}}]{finite40}%
  \BibitemOpen
  \bibfield  {author} {\bibinfo {author} {\bibfnamefont {J.}~\bibnamefont
  {Braun}}, \bibinfo {author} {\bibfnamefont {B.}~\bibnamefont {Klein}}, \ and\
  \bibinfo {author} {\bibfnamefont {H.~J.}\ \bibnamefont {Pirner}},\ }\href
  {\doibase 10.1103/PhysRevD.71.014032} {\bibfield  {journal} {\bibinfo
  {journal} {Phys. Rev. D}\ }\textbf {\bibinfo {volume} {71}},\ \bibinfo
  {pages} {014032} (\bibinfo {year} {2005}{\natexlab{a}})}\BibitemShut
  {NoStop}%
\bibitem [{\citenamefont {Braun}\ \emph
  {et~al.}(2005{\natexlab{b}})\citenamefont {Braun}, \citenamefont {Klein},\
  and\ \citenamefont {Pirner}}]{finite41}%
  \BibitemOpen
  \bibfield  {author} {\bibinfo {author} {\bibfnamefont {J.}~\bibnamefont
  {Braun}}, \bibinfo {author} {\bibfnamefont {B.}~\bibnamefont {Klein}}, \ and\
  \bibinfo {author} {\bibfnamefont {H.~J.}\ \bibnamefont {Pirner}},\ }\href
  {\doibase 10.1103/PhysRevD.72.034017} {\bibfield  {journal} {\bibinfo
  {journal} {Phys. Rev. D}\ }\textbf {\bibinfo {volume} {72}},\ \bibinfo
  {pages} {034017} (\bibinfo {year} {2005}{\natexlab{b}})}\BibitemShut
  {NoStop}%
\bibitem [{\citenamefont {Braun}\ \emph {et~al.}(2006)\citenamefont {Braun},
  \citenamefont {Klein}, \citenamefont {Pirner},\ and\ \citenamefont
  {Rezaeian}}]{finite42}%
  \BibitemOpen
  \bibfield  {author} {\bibinfo {author} {\bibfnamefont {J.}~\bibnamefont
  {Braun}}, \bibinfo {author} {\bibfnamefont {B.}~\bibnamefont {Klein}},
  \bibinfo {author} {\bibfnamefont {H.-J.}\ \bibnamefont {Pirner}}, \ and\
  \bibinfo {author} {\bibfnamefont {A.~H.}\ \bibnamefont {Rezaeian}},\ }\href
  {\doibase 10.1103/PhysRevD.73.074010} {\bibfield  {journal} {\bibinfo
  {journal} {Phys. Rev. D}\ }\textbf {\bibinfo {volume} {73}},\ \bibinfo
  {pages} {074010} (\bibinfo {year} {2006})}\BibitemShut {NoStop}%
\bibitem [{\citenamefont {Yamamoto}\ and\ \citenamefont
  {Kanazawa}(2009)}]{finite5}%
  \BibitemOpen
  \bibfield  {author} {\bibinfo {author} {\bibfnamefont {N.}~\bibnamefont
  {Yamamoto}}\ and\ \bibinfo {author} {\bibfnamefont {T.}~\bibnamefont
  {Kanazawa}},\ }\href {\doibase 10.1103/PhysRevLett.103.032001} {\bibfield
  {journal} {\bibinfo  {journal} {Phys. Rev. Lett.}\ }\textbf {\bibinfo
  {volume} {103}},\ \bibinfo {pages} {032001} (\bibinfo {year}
  {2009})}\BibitemShut {NoStop}%
\bibitem [{\citenamefont {Palhares}\ \emph {et~al.}(2010)\citenamefont
  {Palhares}, \citenamefont {Fraga},\ and\ \citenamefont {Kodama}}]{finite6}%
  \BibitemOpen
  \bibfield  {author} {\bibinfo {author} {\bibfnamefont {L.~F.}\ \bibnamefont
  {Palhares}}, \bibinfo {author} {\bibfnamefont {E.~S.}\ \bibnamefont {Fraga}},
  \ and\ \bibinfo {author} {\bibfnamefont {T.}~\bibnamefont {Kodama}},\ }\href
  {\doibase 10.1088/0954-3899/37/9/094031} {\bibfield  {journal} {\bibinfo
  {journal} {J. Phys. G: Nucl. Part. Phys.}\ }\textbf {\bibinfo {volume}
  {37}},\ \bibinfo {pages} {094031} (\bibinfo {year} {2010})}\BibitemShut
  {NoStop}%
\bibitem [{\citenamefont {Spieles}\ \emph {et~al.}(1998)\citenamefont
  {Spieles}, \citenamefont {St\"ocker},\ and\ \citenamefont
  {Greiner}}]{finite7}%
  \BibitemOpen
  \bibfield  {author} {\bibinfo {author} {\bibfnamefont {C.}~\bibnamefont
  {Spieles}}, \bibinfo {author} {\bibfnamefont {H.}~\bibnamefont {St\"ocker}},
  \ and\ \bibinfo {author} {\bibfnamefont {C.}~\bibnamefont {Greiner}},\ }\href
  {\doibase 10.1103/PhysRevC.57.908} {\bibfield  {journal} {\bibinfo  {journal}
  {Phys. Rev. C}\ }\textbf {\bibinfo {volume} {57}},\ \bibinfo {pages} {908}
  (\bibinfo {year} {1998})}\BibitemShut {NoStop}%
\bibitem [{\citenamefont {Fraga}\ and\ \citenamefont
  {Venugopalan}(2005)}]{finite8}%
  \BibitemOpen
  \bibfield  {author} {\bibinfo {author} {\bibfnamefont {E.~S.}\ \bibnamefont
  {Fraga}}\ and\ \bibinfo {author} {\bibfnamefont {R.}~\bibnamefont
  {Venugopalan}},\ }\href {\doibase
  https://doi.org/10.1016/j.physa.2004.07.045} {\bibfield  {journal} {\bibinfo
  {journal} {Physica A}\ }\textbf {\bibinfo {volume} {345}},\ \bibinfo {pages}
  {121 } (\bibinfo {year} {2005})}\BibitemShut {NoStop}%
\bibitem [{\citenamefont {Hansen}(1990)}]{chiral2}%
  \BibitemOpen
  \bibfield  {author} {\bibinfo {author} {\bibfnamefont {F.}~\bibnamefont
  {Hansen}},\ }\href {\doibase https://doi.org/10.1016/0550-3213(90)90405-3}
  {\bibfield  {journal} {\bibinfo  {journal} {Nucl. Phys. B}\ }\textbf
  {\bibinfo {volume} {345}},\ \bibinfo {pages} {685 } (\bibinfo {year}
  {1990})}\BibitemShut {NoStop}%
\bibitem [{\citenamefont {Damgaard}\ and\ \citenamefont
  {Fukaya}(2009)}]{chiral3}%
  \BibitemOpen
  \bibfield  {author} {\bibinfo {author} {\bibfnamefont {P.~H.}\ \bibnamefont
  {Damgaard}}\ and\ \bibinfo {author} {\bibfnamefont {H.}~\bibnamefont
  {Fukaya}},\ }\href {\doibase 10.1088/1126-6708/2009/01/052} {\bibfield
  {journal} {\bibinfo  {journal} {J. High Energy Phys.}\ }\textbf {\bibinfo
  {volume} {2009}},\ \bibinfo {pages} {052} (\bibinfo {year}
  {2009})}\BibitemShut {NoStop}%
\bibitem [{\citenamefont {Colangelo}\ \emph {et~al.}(2005)\citenamefont
  {Colangelo}, \citenamefont {Dürr},\ and\ \citenamefont
  {Haefeli}}]{quark-meson3}%
  \BibitemOpen
  \bibfield  {author} {\bibinfo {author} {\bibfnamefont {G.}~\bibnamefont
  {Colangelo}}, \bibinfo {author} {\bibfnamefont {S.}~\bibnamefont {Dürr}}, \
  and\ \bibinfo {author} {\bibfnamefont {C.}~\bibnamefont {Haefeli}},\ }\href
  {\doibase https://doi.org/10.1016/j.nuclphysb.2005.05.015} {\bibfield
  {journal} {\bibinfo  {journal} {Nucl. Phys. B}\ }\textbf {\bibinfo {volume}
  {721}},\ \bibinfo {pages} {136 } (\bibinfo {year} {2005})}\BibitemShut
  {NoStop}%
\bibitem [{\citenamefont {Colangelo}\ \emph {et~al.}(2010)\citenamefont
  {Colangelo}, \citenamefont {Fuhrer},\ and\ \citenamefont
  {Lanz}}]{quark-meson4}%
  \BibitemOpen
  \bibfield  {author} {\bibinfo {author} {\bibfnamefont {G.}~\bibnamefont
  {Colangelo}}, \bibinfo {author} {\bibfnamefont {A.}~\bibnamefont {Fuhrer}}, \
  and\ \bibinfo {author} {\bibfnamefont {S.}~\bibnamefont {Lanz}},\ }\href
  {\doibase 10.1103/PhysRevD.82.034506} {\bibfield  {journal} {\bibinfo
  {journal} {Phys. Rev. D}\ }\textbf {\bibinfo {volume} {82}},\ \bibinfo
  {pages} {034506} (\bibinfo {year} {2010})}\BibitemShut {NoStop}%
\bibitem [{\citenamefont {Luecker}\ \emph {et~al.}(2010)\citenamefont
  {Luecker}, \citenamefont {Fischer},\ and\ \citenamefont {Williams}}]{DS1}%
  \BibitemOpen
  \bibfield  {author} {\bibinfo {author} {\bibfnamefont {J.}~\bibnamefont
  {Luecker}}, \bibinfo {author} {\bibfnamefont {C.~S.}\ \bibnamefont
  {Fischer}}, \ and\ \bibinfo {author} {\bibfnamefont {R.}~\bibnamefont
  {Williams}},\ }\href {\doibase 10.1103/PhysRevD.81.094005} {\bibfield
  {journal} {\bibinfo  {journal} {Phys. Rev. D}\ }\textbf {\bibinfo {volume}
  {81}},\ \bibinfo {pages} {094005} (\bibinfo {year} {2010})}\BibitemShut
  {NoStop}%
\bibitem [{\citenamefont {Li}\ \emph {et~al.}(2019)\citenamefont {Li},
  \citenamefont {Cui}, \citenamefont {Zhou}, \citenamefont {An}, \citenamefont
  {Zhang},\ and\ \citenamefont {Zong}}]{DS2}%
  \BibitemOpen
  \bibfield  {author} {\bibinfo {author} {\bibfnamefont {B.-L.}\ \bibnamefont
  {Li}}, \bibinfo {author} {\bibfnamefont {Z.-F.}\ \bibnamefont {Cui}},
  \bibinfo {author} {\bibfnamefont {B.-W.}\ \bibnamefont {Zhou}}, \bibinfo
  {author} {\bibfnamefont {S.}~\bibnamefont {An}}, \bibinfo {author}
  {\bibfnamefont {L.-P.}\ \bibnamefont {Zhang}}, \ and\ \bibinfo {author}
  {\bibfnamefont {H.-S.}\ \bibnamefont {Zong}},\ }\href {\doibase
  https://doi.org/10.1016/j.nuclphysb.2018.11.015} {\bibfield  {journal}
  {\bibinfo  {journal} {Nucl. Phys. B}\ }\textbf {\bibinfo {volume} {938}},\
  \bibinfo {pages} {298 } (\bibinfo {year} {2019})}\BibitemShut {NoStop}%
\bibitem [{\citenamefont {Shi}\ \emph {et~al.}(2018{\natexlab{a}})\citenamefont
  {Shi}, \citenamefont {Jia}, \citenamefont {Sun}, \citenamefont {Zhang},\ and\
  \citenamefont {Zong}}]{DS3}%
  \BibitemOpen
  \bibfield  {author} {\bibinfo {author} {\bibfnamefont {C.}~\bibnamefont
  {Shi}}, \bibinfo {author} {\bibfnamefont {W.-B.}\ \bibnamefont {Jia}},
  \bibinfo {author} {\bibfnamefont {A.}~\bibnamefont {Sun}}, \bibinfo {author}
  {\bibfnamefont {L.-P.}\ \bibnamefont {Zhang}}, \ and\ \bibinfo {author}
  {\bibfnamefont {H.-S.}\ \bibnamefont {Zong}},\ }\href {\doibase
  10.1088/1674-1137/42/2/023101} {\bibfield  {journal} {\bibinfo  {journal}
  {Chin. Phys. C}\ }\textbf {\bibinfo {volume} {42}},\ \bibinfo {pages}
  {023101} (\bibinfo {year} {2018}{\natexlab{a}})}\BibitemShut {NoStop}%
\bibitem [{\citenamefont {Shi}\ \emph {et~al.}(2018{\natexlab{b}})\citenamefont
  {Shi}, \citenamefont {Xia}, \citenamefont {Jia},\ and\ \citenamefont
  {Zong}}]{DS4}%
  \BibitemOpen
  \bibfield  {author} {\bibinfo {author} {\bibfnamefont {C.}~\bibnamefont
  {Shi}}, \bibinfo {author} {\bibfnamefont {Y.-H.}\ \bibnamefont {Xia}},
  \bibinfo {author} {\bibfnamefont {W.-B.}\ \bibnamefont {Jia}}, \ and\
  \bibinfo {author} {\bibfnamefont {H.-S.}\ \bibnamefont {Zong}},\ }\href
  {\doibase 10.1007/s11433-017-9177-4} {\bibfield  {journal} {\bibinfo
  {journal} {Sci. China Phys, Mech.}\ }\textbf {\bibinfo {volume} {61}},\
  \bibinfo {pages} {082021} (\bibinfo {year} {2018}{\natexlab{b}})}\BibitemShut
  {NoStop}%
\bibitem [{\citenamefont {Bhattacharyya}\ \emph {et~al.}(2015)\citenamefont
  {Bhattacharyya}, \citenamefont {Ray},\ and\ \citenamefont {Sur}}]{pNJL1}%
  \BibitemOpen
  \bibfield  {author} {\bibinfo {author} {\bibfnamefont {A.}~\bibnamefont
  {Bhattacharyya}}, \bibinfo {author} {\bibfnamefont {R.}~\bibnamefont {Ray}},
  \ and\ \bibinfo {author} {\bibfnamefont {S.}~\bibnamefont {Sur}},\ }\href
  {\doibase 10.1103/PhysRevD.91.051501} {\bibfield  {journal} {\bibinfo
  {journal} {Phys. Rev. D}\ }\textbf {\bibinfo {volume} {91}},\ \bibinfo
  {pages} {051501} (\bibinfo {year} {2015})}\BibitemShut {NoStop}%
\bibitem [{\citenamefont {Bhattacharyya}\ \emph {et~al.}(2013)\citenamefont
  {Bhattacharyya}, \citenamefont {Deb}, \citenamefont {Ghosh}, \citenamefont
  {Ray},\ and\ \citenamefont {Sur}}]{pNJL2}%
  \BibitemOpen
  \bibfield  {author} {\bibinfo {author} {\bibfnamefont {A.}~\bibnamefont
  {Bhattacharyya}}, \bibinfo {author} {\bibfnamefont {P.}~\bibnamefont {Deb}},
  \bibinfo {author} {\bibfnamefont {S.~K.}\ \bibnamefont {Ghosh}}, \bibinfo
  {author} {\bibfnamefont {R.}~\bibnamefont {Ray}}, \ and\ \bibinfo {author}
  {\bibfnamefont {S.}~\bibnamefont {Sur}},\ }\href {\doibase
  10.1103/PhysRevD.87.054009} {\bibfield  {journal} {\bibinfo  {journal} {Phys.
  Rev. D}\ }\textbf {\bibinfo {volume} {87}},\ \bibinfo {pages} {054009}
  (\bibinfo {year} {2013})}\BibitemShut {NoStop}%
\bibitem [{\citenamefont {Pan}\ \emph {et~al.}(2017)\citenamefont {Pan},
  \citenamefont {Cui}, \citenamefont {Chang},\ and\ \citenamefont
  {Zong}}]{pNJL3}%
  \BibitemOpen
  \bibfield  {author} {\bibinfo {author} {\bibfnamefont {Z.}~\bibnamefont
  {Pan}}, \bibinfo {author} {\bibfnamefont {Z.-F.}\ \bibnamefont {Cui}},
  \bibinfo {author} {\bibfnamefont {C.-H.}\ \bibnamefont {Chang}}, \ and\
  \bibinfo {author} {\bibfnamefont {H.-S.}\ \bibnamefont {Zong}},\ }\href
  {\doibase 10.1142/S0217751X17500671} {\bibfield  {journal} {\bibinfo
  {journal} {Int. J. Phys. A}\ }\textbf {\bibinfo {volume} {32}},\ \bibinfo
  {pages} {1750067} (\bibinfo {year} {2017})}\BibitemShut {NoStop}%
\bibitem [{\citenamefont {Tripolt}\ \emph {et~al.}(2014)\citenamefont
  {Tripolt}, \citenamefont {Braun}, \citenamefont {Klein},\ and\ \citenamefont
  {Schaefer}}]{fluctuation}%
  \BibitemOpen
  \bibfield  {author} {\bibinfo {author} {\bibfnamefont {R.-A.}\ \bibnamefont
  {Tripolt}}, \bibinfo {author} {\bibfnamefont {J.}~\bibnamefont {Braun}},
  \bibinfo {author} {\bibfnamefont {B.}~\bibnamefont {Klein}}, \ and\ \bibinfo
  {author} {\bibfnamefont {B.-J.}\ \bibnamefont {Schaefer}},\ }\href {\doibase
  10.1103/PhysRevD.90.054012} {\bibfield  {journal} {\bibinfo  {journal} {Phys.
  Rev. D}\ }\textbf {\bibinfo {volume} {90}},\ \bibinfo {pages} {054012}
  (\bibinfo {year} {2014})}\BibitemShut {NoStop}%
\bibitem [{\citenamefont {Klein}(2017)}]{review}%
  \BibitemOpen
  \bibfield  {author} {\bibinfo {author} {\bibfnamefont {B.}~\bibnamefont
  {Klein}},\ }\href {\doibase https://doi.org/10.1016/j.physrep.2017.09.002}
  {\bibfield  {journal} {\bibinfo  {journal} {Phys. Rep.}\ }\textbf {\bibinfo
  {volume} {707-708}},\ \bibinfo {pages} {1 } (\bibinfo {year}
  {2017})}\BibitemShut {NoStop}%
\bibitem [{\citenamefont {Kiriyama}(2005)}]{MRE2}%
  \BibitemOpen
  \bibfield  {author} {\bibinfo {author} {\bibfnamefont {O.}~\bibnamefont
  {Kiriyama}},\ }\href {\doibase 10.1103/PhysRevD.72.054009} {\bibfield
  {journal} {\bibinfo  {journal} {Phys. Rev. D}\ }\textbf {\bibinfo {volume}
  {72}},\ \bibinfo {pages} {054009} (\bibinfo {year} {2005})}\BibitemShut
  {NoStop}%
\bibitem [{\citenamefont {Lugones}\ \emph {et~al.}(2013)\citenamefont
  {Lugones}, \citenamefont {Grunfeld},\ and\ \citenamefont {Ajmi}}]{MRE3}%
  \BibitemOpen
  \bibfield  {author} {\bibinfo {author} {\bibfnamefont {G.}~\bibnamefont
  {Lugones}}, \bibinfo {author} {\bibfnamefont {A.~G.}\ \bibnamefont
  {Grunfeld}}, \ and\ \bibinfo {author} {\bibfnamefont {M.~A.}\ \bibnamefont
  {Ajmi}},\ }\href {\doibase 10.1103/PhysRevC.88.045803} {\bibfield  {journal}
  {\bibinfo  {journal} {Phys. Rev. C}\ }\textbf {\bibinfo {volume} {88}},\
  \bibinfo {pages} {045803} (\bibinfo {year} {2013})}\BibitemShut {NoStop}%
\bibitem [{\citenamefont {Zhao}\ \emph {et~al.}(2019)\citenamefont {Zhao},
  \citenamefont {Zhang}, \citenamefont {Zhang},\ and\ \citenamefont
  {Zong}}]{MRE4}%
  \BibitemOpen
  \bibfield  {author} {\bibinfo {author} {\bibfnamefont {Y.-P.}\ \bibnamefont
  {Zhao}}, \bibinfo {author} {\bibfnamefont {R.-R.}\ \bibnamefont {Zhang}},
  \bibinfo {author} {\bibfnamefont {H.}~\bibnamefont {Zhang}}, \ and\ \bibinfo
  {author} {\bibfnamefont {H.-S.}\ \bibnamefont {Zong}},\ }\href {\doibase
  10.1088/1674-1137/43/6/063101} {\bibfield  {journal} {\bibinfo  {journal}
  {Chin. Phys. C}\ }\textbf {\bibinfo {volume} {43}},\ \bibinfo {pages}
  {063101} (\bibinfo {year} {2019})}\BibitemShut {NoStop}%
\bibitem [{\citenamefont {Balian}\ and\ \citenamefont {Bloch}(1970)}]{mre}%
  \BibitemOpen
  \bibfield  {author} {\bibinfo {author} {\bibfnamefont {R.}~\bibnamefont
  {Balian}}\ and\ \bibinfo {author} {\bibfnamefont {C.}~\bibnamefont {Bloch}},\
  }\href {\doibase https://doi.org/10.1016/0003-4916(70)90497-5} {\bibfield
  {journal} {\bibinfo  {journal} {Ann. Phys.}\ }\textbf {\bibinfo {volume}
  {60}},\ \bibinfo {pages} {401 } (\bibinfo {year} {1970})}\BibitemShut
  {NoStop}%
\bibitem [{\citenamefont {Klevansky}(1992)}]{NJLreview}%
  \BibitemOpen
  \bibfield  {author} {\bibinfo {author} {\bibfnamefont {S.~P.}\ \bibnamefont
  {Klevansky}},\ }\href {\doibase 10.1103/RevModPhys.64.649} {\bibfield
  {journal} {\bibinfo  {journal} {Rev. Mod. Phys.}\ }\textbf {\bibinfo {volume}
  {64}},\ \bibinfo {pages} {649} (\bibinfo {year} {1992})}\BibitemShut
  {NoStop}%
\bibitem [{\citenamefont {Wang}\ \emph {et~al.}(2018)\citenamefont {Wang},
  \citenamefont {Xia},\ and\ \citenamefont {Zong}}]{qingwu}%
  \BibitemOpen
  \bibfield  {author} {\bibinfo {author} {\bibfnamefont {Q.-W.}\ \bibnamefont
  {Wang}}, \bibinfo {author} {\bibfnamefont {Y.-H.}\ \bibnamefont {Xia}}, \
  and\ \bibinfo {author} {\bibfnamefont {H.-S.}\ \bibnamefont {Zong}},\ }\href
  {\doibase 10.1142/S0217732318502322} {\bibfield  {journal} {\bibinfo
  {journal} {Mod. Phys. Lett. A}\ }\textbf {\bibinfo {volume} {33}},\ \bibinfo
  {pages} {1850232} (\bibinfo {year} {2018})}\BibitemShut {NoStop}%
\bibitem [{\citenamefont {Xia}\ \emph {et~al.}(2019)\citenamefont {Xia},
  \citenamefont {Wang}, \citenamefont {Feng},\ and\ \citenamefont
  {Zong}}]{Xia}%
  \BibitemOpen
  \bibfield  {author} {\bibinfo {author} {\bibfnamefont {Y.-H.}\ \bibnamefont
  {Xia}}, \bibinfo {author} {\bibfnamefont {Q.-W.}\ \bibnamefont {Wang}},
  \bibinfo {author} {\bibfnamefont {H.-T.}\ \bibnamefont {Feng}}, \ and\
  \bibinfo {author} {\bibfnamefont {H.-S.}\ \bibnamefont {Zong}},\ }\href
  {\doibase 10.1088/1674-1137/43/3/034101} {\bibfield  {journal} {\bibinfo
  {journal} {Chin. Phys. C}\ }\textbf {\bibinfo {volume} {43}},\ \bibinfo
  {pages} {034101} (\bibinfo {year} {2019})}\BibitemShut {NoStop}%
\bibitem [{\citenamefont {Alonso}\ \emph {et~al.}(1997)\citenamefont {Alonso},
  \citenamefont {Vincenzo},\ and\ \citenamefont {Mondino}}]{Diraceq}%
  \BibitemOpen
  \bibfield  {author} {\bibinfo {author} {\bibfnamefont {V.}~\bibnamefont
  {Alonso}}, \bibinfo {author} {\bibfnamefont {S.~D.}\ \bibnamefont
  {Vincenzo}}, \ and\ \bibinfo {author} {\bibfnamefont {L.}~\bibnamefont
  {Mondino}},\ }\href {\doibase 10.1088/0143-0807/18/5/001} {\bibfield
  {journal} {\bibinfo  {journal} {Eur. J. Phys.}\ }\textbf {\bibinfo {volume}
  {18}},\ \bibinfo {pages} {315} (\bibinfo {year} {1997})}\BibitemShut
  {NoStop}%
\bibitem [{\citenamefont {Chodos}\ \emph
  {et~al.}(1974{\natexlab{a}})\citenamefont {Chodos}, \citenamefont {Jaffe},
  \citenamefont {Johnson}, \citenamefont {Thorn},\ and\ \citenamefont
  {Weisskopf}}]{MIT1}%
  \BibitemOpen
  \bibfield  {author} {\bibinfo {author} {\bibfnamefont {A.}~\bibnamefont
  {Chodos}}, \bibinfo {author} {\bibfnamefont {R.~L.}\ \bibnamefont {Jaffe}},
  \bibinfo {author} {\bibfnamefont {K.}~\bibnamefont {Johnson}}, \bibinfo
  {author} {\bibfnamefont {C.~B.}\ \bibnamefont {Thorn}}, \ and\ \bibinfo
  {author} {\bibfnamefont {V.~F.}\ \bibnamefont {Weisskopf}},\ }\href {\doibase
  10.1103/PhysRevD.9.3471} {\bibfield  {journal} {\bibinfo  {journal} {Phys.
  Rev. D}\ }\textbf {\bibinfo {volume} {9}},\ \bibinfo {pages} {3471} (\bibinfo
  {year} {1974}{\natexlab{a}})}\BibitemShut {NoStop}%
\bibitem [{\citenamefont {Chodos}\ \emph
  {et~al.}(1974{\natexlab{b}})\citenamefont {Chodos}, \citenamefont {Jaffe},
  \citenamefont {Johnson},\ and\ \citenamefont {Thorn}}]{MIT2}%
  \BibitemOpen
  \bibfield  {author} {\bibinfo {author} {\bibfnamefont {A.}~\bibnamefont
  {Chodos}}, \bibinfo {author} {\bibfnamefont {R.~L.}\ \bibnamefont {Jaffe}},
  \bibinfo {author} {\bibfnamefont {K.}~\bibnamefont {Johnson}}, \ and\
  \bibinfo {author} {\bibfnamefont {C.~B.}\ \bibnamefont {Thorn}},\ }\href
  {\doibase 10.1103/PhysRevD.10.2599} {\bibfield  {journal} {\bibinfo
  {journal} {Phys. Rev. D}\ }\textbf {\bibinfo {volume} {10}},\ \bibinfo
  {pages} {2599} (\bibinfo {year} {1974}{\natexlab{b}})}\BibitemShut {NoStop}%
\bibitem [{\citenamefont {Chernodub}\ and\ \citenamefont
  {Gongyo}(2017)}]{cylinder}%
  \BibitemOpen
  \bibfield  {author} {\bibinfo {author} {\bibfnamefont {M.~N.}\ \bibnamefont
  {Chernodub}}\ and\ \bibinfo {author} {\bibfnamefont {S.}~\bibnamefont
  {Gongyo}},\ }\href {\doibase 10.1007/JHEP01(2017)136} {\bibfield  {journal}
  {\bibinfo  {journal} {J. High Energy Phys.}\ }\textbf {\bibinfo {volume}
  {2017}},\ \bibinfo {pages} {136} (\bibinfo {year} {2017})}\BibitemShut
  {NoStop}%
\bibitem [{\citenamefont {{Greiner}}\ \emph {et~al.}(2007)\citenamefont
  {{Greiner}}, \citenamefont {{Schramm}},\ and\ \citenamefont
  {{Stein}}}]{Greiner}%
  \BibitemOpen
  \bibfield  {author} {\bibinfo {author} {\bibfnamefont {W.}~\bibnamefont
  {{Greiner}}}, \bibinfo {author} {\bibfnamefont {S.}~\bibnamefont
  {{Schramm}}}, \ and\ \bibinfo {author} {\bibfnamefont {E.}~\bibnamefont
  {{Stein}}},\ }\href@noop {} {\emph {\bibinfo {title} {Quantum Chromodynamics,
  Third edition}}}\ (\bibinfo  {publisher} {Springer-Verlag},\ \bibinfo {year}
  {2007})\BibitemShut {NoStop}%
\bibitem [{\citenamefont {Leutwyler}\ and\ \citenamefont
  {Smilga}(1992)}]{Leut}%
  \BibitemOpen
  \bibfield  {author} {\bibinfo {author} {\bibfnamefont {H.}~\bibnamefont
  {Leutwyler}}\ and\ \bibinfo {author} {\bibfnamefont {A.}~\bibnamefont
  {Smilga}},\ }\href {\doibase 10.1103/PhysRevD.46.5607} {\bibfield  {journal}
  {\bibinfo  {journal} {Phys. Rev. D}\ }\textbf {\bibinfo {volume} {46}},\
  \bibinfo {pages} {5607} (\bibinfo {year} {1992})}\BibitemShut {NoStop}%
\bibitem [{\citenamefont {Weinberg}(2013)}]{Weinberg:1996kr}%
  \BibitemOpen
  \bibfield  {author} {\bibinfo {author} {\bibfnamefont {S.}~\bibnamefont
  {Weinberg}},\ }\href@noop {} {\emph {\bibinfo {title} {{The quantum theory of
  fields. Vol. 2: Modern applications}}}}\ (\bibinfo  {publisher} {Cambridge
  University Press},\ \bibinfo {year} {2013})\BibitemShut {NoStop}%
\end{thebibliography}%
\end{document}